\documentclass[11pt]{article}
\usepackage[a4paper]{geometry}
\usepackage{clic2017}
\usepackage{times}
\usepackage{url}
\usepackage{latexsym}
\usepackage{graphicx}

\title{Multilingual, Temporal and Sentimental Distant-Reading of City Events}

\author{Mehmet Can Yavuz \\
Faculty of Engineering and Natural Science, Sabancı University, Tuzla\\
İstanbul, Türkiye\\
{\tt mehmetyavuz@sabanciuniv.edu}}

\date{}

\begin{document}
\maketitle
\begin{abstract}
Leibniz’s Monadology mentions perceptional and sentimental variations of the individual in the city. It is the interaction of people with people and events. Film festivals are highly sentimental events of multicultural cities. Each movie has a different sentimental effect and the interactions with the movies have reflections that can be observed on social media. This analysis aims to apply distant reading on Berlinale tweets collected during the festival. On contrary to close reading, distant reading let authors to observe patterns in large collection of data. The analysis is temporal and sentimental in multilingual domain and strongly positive and negative time intervals are analysed.

For this purpose, we trained a deep sentiment network with multilingual embeddings. These multilingual embeddings are aligned in latent space. We trained the network with a multilingual dataset in three languages English, German and Spanish. The trained algorithm has a 0.78 test score and applied on Tweets with Berlinale hashtag during the festival.

Although the sentimental analysis does not reflect the award-winning films, we observe weekly routine on the relationship between sentimentality, which can mislead a close reading analysis. We have also remarks on popularity of the director or actors.
\end{abstract}

\section{Introduction}
{\let\thefootnote\relax\footnote{{Copyright © 2020 for this paper by its authors. Use permitted under Creative Commons License Attribution 4.0 International (CC BY 4.0).}}} 
When Leibniz wrote Monadology in the 18th century, he also mentioned perceptual change of monad. For example, monad or the individual is constantly interacting in the city and is subject to perceptual and sentimental changes. He saw someone happy, he saw someone else and his mood was broken ... or he watched a movie in a theatre. Film festivals constitute one of the most important events of modern cities. People, or, as Leibniz puts it, monads follow the movies competing for that year's grand prize. On the last day of the festivals, there is a reward ceremony. It can be assumed that there is a difference between the jury members who give great awards and the audience's taste. One of the primary ways to observe this interaction is to use social media. By looking at the emotional analysis of the festival tweets published with the Berlinale hashtag, such as in this study, we can have an observation on events and the city life.

The problem with the close reading of tweets, is that, the patterns may not be observable. The distant reading of such datasets can lead author to conclude more reliable on their analyses. The term is coined by Franco Moretti to describe the process of locating patterns across a large corpus of text. Distance, for Moretti, offers 'a specific form of knowledge: fewer elements, hence a sharper sense of their overall interconnection' (Moretti, 2005). Here, we perform a distant reading of audience-generated texts, this article offers an important observation of such dataset on events and city life. For this purpose, we collected the 2019 tweets of the Berlin Film Festival and made an analysis on the audience-festival interaction. The tweets published with the Berlinale hashtag are basically in three languages. These tweets, which are in English, German and Spanish, have been subjected to multilingual sentiment analysis. When we sort the tweets according to their release date, the temporal feature of the analysis is revealed. Thus, the analysis is a multilingual, temporal and sentimental analysis. Subjecting sentimental scores to outlier detection analysis can reveal intervals of intense likes and dislikes. We are concerned with the analysis of these intervals.

\subsection{Related Works}
Sentiment analysis of social media is a widely researched topic, (Ortigosa, 2014; Pak, 2004). Here, we would like to do a multilingual sentiment analysis only by using Recurrent Neural Networks. Unlike our approach, there are lexicon based approaches, (Taboada, 2011; Banea, 2008). There are also machine translation based methods. Hiroshi et al.(Kanayama, 2004) translated only sentiment units. Balahur and Turchi (Balahur, 2014) used uni-grams, bi-grams and tf-idf features. There is also literature that uses deep learning models trained on one language and the translations are processed (Can, 2018). 

In this research, the generalized (extreme Studentized deviate) ESD test helps exploring outliers in a univariate data set that follows an approximately normal distribution (Rosner, 1983). Also, multilingual word embeddings (MUSE) are used for word representations among three languages (Conneau, 2017), these embeddings are aligned in semantic space.

\section{Methodology}
Our analyses are based on three basic elements: MUSE, sentiment network, ESD. The first one is to use multilingual embeddings, since analyses are in three languages. The vectors of similar words in three languages are parallel to each other in semantic space. This item is for making optimal convergence for LSTM sentiment analysis. By this way, the network can easily be converged. The sentimental score of each tweet was determined and anomalies were detected by Extreme Studentized Deviate (ESD) analysis. Anomalies are meant to identify intensive positive or negative ranges. Apart from these ranges, intense negative tweets and the appearance of the festival were also examined. In the following sections, the theoretical background of these analyzes will be examined respectively. Firstly embeddings and then sentiment networks and lastly ESD analysis are discussed.

\begin{figure}
\includegraphics[width=0.5\textwidth]{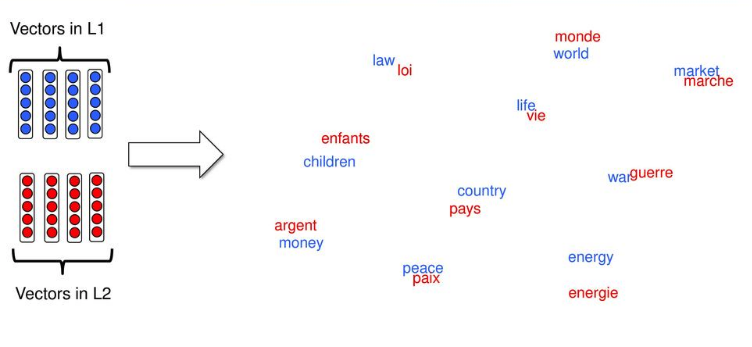}
\caption{The aligned embeddings of different languages according to MUSE embedding alignment.}
\label{fig1}
\end{figure}

\subsection{Multilingual Unsupervised and Supervised Embeddings (Conneau, 2017)}
Embeddings are the pretraining for NLP tasks in Neural Networks. Dimension reduction is applied on the one-hot embeddings of each word. For a vector of length dictionary, output is a fixed length real vector. MUSE embeddings  are multilingual embeddings. For different languages, monolingual embeddings are trained. Then for each language, embeddings are aligned by a linear transformation. By using linear transformation, semantic space of different languages intersects. For example, the word "cat" would be located in similar locations.

Figure 1 clearly shows the process of alignment. (A) At the beginning, there are two different monolingual manifolds. Let's say English words are red and Italian words are blue. In this process, one of the manifolds would be transformed into the semantic space of the other. (B) A rotation matrix W is learnt by using adversarial learning. The green stars are randomly selected words for discriminator to determine whether the two word embeddings come from the same distribution. (C) The mapping is improved by using alignment of frequent words. (D) Finally, the mapping W translates and a distance metric expands the space for high density of points. By distance metric, the words around "cat" expands, and becomes similar to the region around the Italian "gatto".

By using multilingual embeddings, it is easier to converge for multilingual training of a network. Since the similar words are aligned in similar locations, the decision boundaries are easier to converge.

\subsection{LSTM Sentiment Network}
Our model inputs the words as MUSE embeddings. Following the pre-trained embedding layer, there is two layer LSTM, followed by an average pooling layer and sigmoid output. 

For the input sequence of embeddings, the memory cells in the LSTM outputs a representation sequence. According to the length of sequence, each sequential representation is averaged. At the top, there is a sigmoid output layer. The output is a binary classification, it is either positive or negative. The output range is between 0 and 1. This architecture is the basic settings for sentiment analysis, the only difference is the fixed pretrained multilingual embeddings.

\subsection{ESD Analysis (Rosner, 1983)}
\begin{figure}[b!]
\centering
\includegraphics[width=0.5\textwidth]{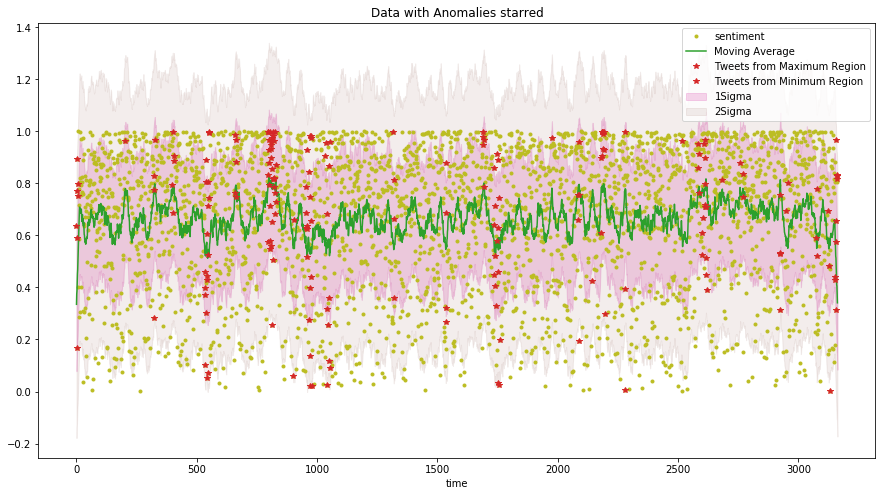}
\caption{The sentimental scores of the tweets ordered in time. The moving average curve and anomalies are indicated.} \label{fig1}
\end{figure}
Aim of the generalized (extreme Studentized deviate) ESD test is to detect one or more outliers in a univariate dataset. The dataset follows an approximately normal distribution. Without specifying the exact number of outliers, the ESD test only requires an upper bound for the number of outliers.

The generalized ESD test performs r separate tests, for a given the upper bound, r. Starting from one outlier, goes on with two outliers and so on. Compute,

\begin{equation}
    R_i = \frac{max_i|x_i-x|}{s}
\end{equation}
with x and s denoting the sample mean and sample standard deviation, respectively.

The observation that maximize $|x_i-x|$ are removed and the above statistics are recomputed with (n-1) observations. Repeat this process until r observations are removed. This results in the r test statistics R1, R2, ..., Rr. Critical values for each statistics are determined by lambda formulation of the paper, then the number of outliers is determined by finding the largest i such that $R_i > \lambda_i$

\section{Experiments}
\begin{figure}[b!]
\centering
\includegraphics[width=0.5\textwidth]{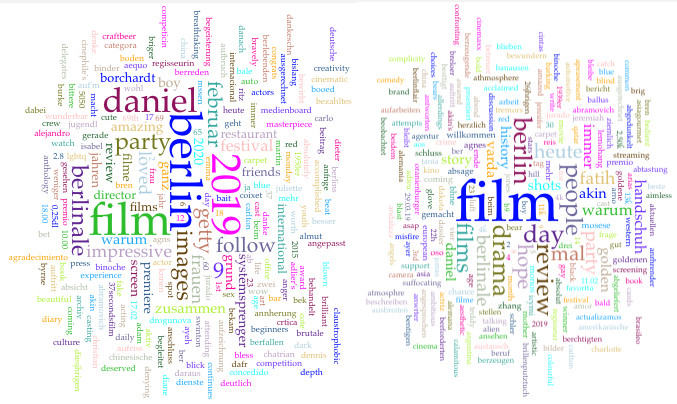}
\caption{Word cloud for negative (left) and positive (right) outlier regions } \label{fig1}
\end{figure}
Training dataset is the subset of Tweets in three different languages, \cite{multilingual}. The available tweets from the annotation table are collected. There are 125K observations in three languages. The dataset is splitted by 80/10/10. These are for the training of a multilingual sentiment network. The data that we analyzed is collected during the festival week, 7-17 February 2019. 3417 tweets that have the Berlinale hashtag are gathered.

LSTM has 300 dimension hidden units, bidirectional and has two layers. The network optimized by 4 epochs and the loss functions are plotted in Figure 2. In the third epoch, it is optimized. The accuracy is 0.78 in multilingual domain. 

\section{Discussions}
Berlinale hashtag related tweets are ordered in time and scored with a multilingual sentiment network. The scores are plotted in Figure 3. The green midline is the moving average with windows size 25. According to ESD analyses, outlier regions are detected. Tweets, that are in abnormally positive or negative regions, are colored with red. There are a couple of observations in these tweets.

The most important observation from the data is on the routine of city life and the festival. The sentimental mood of the festival pretty much depends on the day. The festival is between 7 - 17 February. The first Sunday, 10th of February, is a positive anomaly day. The audience or festival followers are positively outlier. Sentimentally negative outliers only exist on the weekdays. Therefore, if one would like to conclude on a movie solely by using close-reading, the sentimental score can be misleading. The negative sentiments of weekdays or positive sentiments of weekends, the tweets might not be reliable.

In positive outliers, although their frequencies are not high, we observe that famous directors like Fatih Akın and Agres Varda or their films are mentioned. Instead of actors, the festival tends to follow directors. This is in accordance with the general opinion that Berlinale is auteur director oriented festival. In literature, auteur directors are defined as the major creative artists for their films, contrary to industrial cinema (Britannica, 2016).

When we observe the most frequent terms for positive and negative outlier regions, the keywords, "Berlin" and "film" are the most common words. The negative outliers have a high frequency words on the date "2019", "februar" and "9". The positive outliers have the word "drama" and "review" which are related to critics. The negativity is mostly related to films, although the positivity is related to festival related discussion by critics.

\section{Conclusion}
Berlinale is one of the most important multi-cultural events that the reflections upon traced. In this work, we analyzed multi-lingual tweets as time series and find the outlier regions. The results are evaluated in three folds, the directors, the city routines and the frequent terms. We show that Berlinale is auteur director oriented festival as the audience agrees. The city routines are also important at the observations on festival. There is a sentimental mood difference between weekdays and weekends. Lastly, we observe that negativity of the festival is mostly related to the films, unlike positivity. Festival related events are also linked to positive mood.

This analysis is important to show importance of distant-reading on event centric analytic. The analytic mostly depend on close reading of such tweets. The tweets ,which are the interaction of monads or the individual with the city, can be misleading for having an opinion on the films. For such cases, both analysis should be carried out. This is also one explanation for the difference in choice in jury and audience. It is important to mention, the choice of award-winning films for jury is more reliable rather than audience. Audience tends to affect by weekly routine. 

As the film or music festivals are one of the most important events related to city life, it is important to develop further analyses on this track. For this purpose, the multilingual sentiment networks need to be improved as well as more the discussions. For further analysis, it might also be interesting to compare different auteur director oriented festivals over the world. The continental and seasonal differences are also important.


\begin{thebibliography}{00}
\bibitem{mt2}
Balahur, A., Turchi, M. (2014). Comparative experiments using supervised learning and machine translation for multilingual sentiment analysis. \textit{Computer Speech and Language}, 28(1), 56-75.

\bibitem{lex2}
Banea, C., Mihalcea, R., Wiebe, J. (2008, May). A Bootstrapping Method for Building Subjectivity Lexicons for Languages with Scarce Resources. In \textit{LREC} (Vol. 8, pp. 2-764).

\bibitem{autuer}
Britannica (2016). \textit{auteur theory | filmmaking}. [online] Encyclopedia Britannica. Available at: http://www.britannica.com/art/auteur-theory [Accessed 6 Mar. 2016].

\bibitem{mls}
Can, E. F., Ezen-Can, A., Can, F. (2018). Multilingual sentiment analysis: An rnn-based framework for limited data. \textit{arXiv preprint arXiv:1806.04511.}

\bibitem{muse}
Conneau, A., Lample, G., Ranzato, M. A., Denoyer, L., Jégou, H. (2017). Word translation without parallel data. \textit{arXiv preprint arXiv:1710.04087.}

\bibitem{mt1}
Kanayama, H., Nasukawa, T., Watanabe, H. (2004). Deeper sentiment analysis using machine translation technology. In COLING 2004: \textit{Proceedings of the 20th International Conference on Computational Linguistics} (pp. 494-500).

\bibitem{distant}
Moretti, F. (2005). \textit{Graphs, maps, trees: abstract models for a literary history.} Verso.

\bibitem{multilingual}
Mozetič, I., Grčar, M., Smailović, J. (2016). Multilingual Twitter sentiment classification: The role of human annotators. \textit{PloS one}, 11(5).

\bibitem{sent1}
Ortigosa, A., Martín, J. M., Carro, R. M. (2014). Sentiment analysis in Facebook and its application to e-learning. \textit{Computers in human behavior}, 31, 527-541.

\bibitem{sent2}
Pak, A., Paroubek, P. (2010, May). Twitter as a corpus for sentiment analysis and opinion mining. In \textit{LREC} (Vol. 10, No. 2010, pp. 1320-1326).

\bibitem{anomaly}
Rosner, B. (1983). Percentage points for a generalized ESD many-outlier procedure. \textit{Technometrics}, 25(2), 165-172.

\bibitem{lex1}
Taboada, M., Brooke, J., Tofiloski, M., Voll, K.,  Stede, M. (2011). Lexicon-based methods for sentiment analysis. \textit{Computational linguistics}, 37(2), 267-307.

\end{thebibliography}
\end{document}